\documentclass[twocolumn]{aastex62}

\usepackage[nolist]{acronym}

\graphicspath{{./}{figures/}}

\received{September 6, 2018}
\revised{October 1, 2018}
\accepted{October 23, 2018}
\submitjournal{PASP}

\shorttitle{Estimating redshift with k-Nearest Neighbours Regression}
\shortauthors{Luken et al.}

\begin{document}

\title[Estimating redshift with k-Nearest Neighbours Regression]{Preliminary results of using k-Nearest Neighbours Regression to estimate the redshift of radio selected datasets}

\correspondingauthor{Kieran Luken}
\email{k.luken@westernsydney.edu.au}

\author[0000-0002-6147-693X]{Kieran J. Luken}
\affil{Western Sydney University \\
Locked Bag 1797 \\
Penrith, NSW 2751, Australia}
\affil{CSIRO Astronomy and Space Sciences \\
Australia Telescope National Facility \\
PO Box 76, Epping, NSW 1710, Australia
}

\author{Ray P. Norris}
\affil{Western Sydney University \\
Locked Bag 1797 \\
Penrith, NSW 2751, Australia}
\affil{CSIRO Astronomy and Space Sciences \\
Australia Telescope National Facility \\
PO Box 76, Epping, NSW 1710, Australia
}

\author{Laurence A. F. Park}
\affil{Western Sydney University \\
Locked Bag 1797 \\
Penrith, NSW 2751, Australia}



\begin{abstract}

In the near future, all-sky radio surveys are set to produce catalogues of tens of millions of sources with limited multi-wavelength photometry. Spectroscopic redshifts will only be possible for a small fraction of these new-found sources. In this paper, we provide the first in-depth investigation into the use of \ac{kNN} Regression for the estimation of redshift of these sources. We use the \ac{ATLAS} radio data, combined with the \ac{SWIRE} infra-red, the \ac{DES} optical and the \ac{OzDES} spectroscopic survey data. We then reduce the depth of photometry to match what is expected from upcoming \ac{EMU} survey, testing against both data sets. To examine the generalisation of our methods, we test one of the sub-fields of \ac{ATLAS} against the other. We achieve an outlier rate of $\sim$~10\% across all tests, showing that the \ac{kNN} regression algorithm is an acceptable method of estimating redshift, and would perform better given a sample training set with uniform redshift coverage.


\end{abstract}


\section{Introduction} \label{sec:intro}

Large scale radio surveys are becoming more common, resulting in catalogues of millions of radio sources with limited multi-wavelength data \citep{2017NatAs...1..671N}. Knowledge of their redshift  is important to achieve most science goals \citep{EMU}. While spectroscopic redshifts remain the gold standard, only a few million spectroscopic redshifts will be available in this decade with the \ac{SDSS} having measured $\sim$3 million over the northern sky \citep{abolfathi:2018}, the Taipan Galaxy Survey expecting to provide 2 million spectroscopic redshifts out to a redshift of $z= 0.4$ \citep{da_cunha:2017}, and the Wide Area VISTA Extra-galactic Survey (WAVES) expecting to measure 2.5 million redshifts across the souther sky out to a redshift of $z=1.5$ \citep{driver:2016}. Alternatively, redshift can also be measured photometrically, by comparing the magnitudes at different wavelengths to templates \citep{baum, butchins, loh}. Photometric redshifts - or photo-z's - measured using template fitting can be highly accurate, estimating redshift to an accuracy of $ \sigma_{\Delta z / (1 + z_{spec})} \sim 0.015 $ \citep{2011ApJ...742...61S}. However, this requires high quality photometry in at least 15 different filter bands, and up to 31 different bands for the high-accuracy results. Unfortunately, this level of photometry will not be available for large scale sky-surveys. Additionally, photometric template fitting methods tend to fail catastrophically when attempted on AGN, particularly radio-selected AGN \citep{2018MNRAS.473.2655D,salvato_2018}. 

Rather than measuring the redshift directly in the form of spectroscopy, or indirectly by fitting templates, it has been shown that photo-z's can be estimated empirically using the knowledge of previously measured redshifts from similar astronomical objects.

Machine learning has been applied to this problem in the past in the form of neural networks \citep{2003LNCS.2859..226T, 2003MNRAS.339.1195F, 2004PASP..116..345C, 2012A&A...546A..13C, 2013ApJ...772..140B, 2014IAUS..306..307C, 2015ExA....39...45C, 2016PASP..128j4502S, 2017MNRAS.465.1959C, 2018A&A...611A..97P}, random forests \citep{2010ApJ...712..511C, 2015A&A...584A..44C, 2017AA...608A..39M}, the combination of template fitting methods using Bayesian statistics \citep{2018MNRAS.473.2655D} and the stacking of a Self-Organised Map and a Decision Tree \citep{2016MNRAS.460.3152Z}. For the most part, these methods have mainly been concerned with maximising the accuracy of the measured redshift, testing with optically selected galaxy samples, and have been able to achieve a similar accuracy to template fitting methods given a large enough training set. The datasets used have been derived mainly from optical surveys like the \ac{SDSS}, limiting the number of possible radio-loud AGN, which also create issues for photometric template fitting.

\citet{norris:2018} have addressed the problem of relatively low-quality photometry available from all-sky surveys by comparing the performance of different algorithms when using photometry similar to the upcoming \ac{EMU} survey \citep{EMU}, using radio-selected AGN. This has given a glimpse of the expected accuracies of various techniques when the depth of observations and the coverage of all frequency bands is not complete. 

This paper provides a deeper investigation into the effectiveness of the \ac{kNN} algorithm \citep{kNN} for estimating redshift. The \ac{kNN} regression algorithm has previously been applied to photometric redshifts by \citet{2013MNRAS.428..226P}, \citet{2015A&A...576A.132K} and \citet{2017MNRAS.465.1959C}, however \citet{2015A&A...576A.132K} used optical spectra from the Sloan Digital Sky Survey (SDSS), and \citet{2013MNRAS.428..226P} and \citet{2017MNRAS.465.1959C} used spectroscopic redshifts and optical magnitudes only from the SDSS. 
$k$ Nearest Neighbours Regression is a regression model, meaning that it uses a training set of objects and their known redshift values to estimate the redshift of new objects. In particular, kNN regression estimates the redshift of each new object as the mean of the redshifts of the $k$ nearest neighbours from the training set. To perform this task, we must define a feature space (so that we can represent each object as a vector), a metric (to provide the distance between object vectors), and the value of the constant $k$. The feature space dimensions are chosen  as the set of variables that are thought to be predictive of the regression response. Each object is represented as a vector in this space using their measurements of the feature space variables (e.g. in our case each object vectors contain a set of photometry at different wavelengths from a given object in the training set). It is common to find Euclidean distance being used as the kNN metric, and the number of neighbours $k$ typically set within the range of 2 and 20.

Given that the speed of the \ac{kNN} algorithm does not scale well with the number of dimensions or number of sources, this algorithm can be modified to use a k-d tree to find the $k$ most similar sources \citep{2013MNRAS.428..226P}. Alternatively, the problem can be parallelised and run on a GPU. This paper has implemented the \ac{kNN} regression algorithm using the publicly available TensorFlow \footnote{https://www.tensorflow.org/} on GPU \citep{tensorflow2015-whitepaper}, which provides a 3-5 times speed improvement over the equivalent CPU version.

Our initial investigation was to determine the suitability of kNN for the problem. We then tested how well the \ac{kNN} algorithm generalises by testing one sub-field against the other. We next modified our dataset to match the depth expected from the \ac{EMU} survey, and corresponding sky surveys from other astronomical regimes. For all of these tests, we compared the use of Euclidean distance and Manhattan Taxicab distance.

\section{The Data} \label{sec:dataset}

The dataset used in this paper is primarily based on the ATLAS Data Release 3 \citep{ATLAS, franzen}, with cross-identifications to other wavelengths provided by Swan et al. (in preparation). Other catalogues used include the \ac{SWIRE} Infra-red dataset \citep{SWIRE}, the \ac{DES} Optical dataset \citep{DES} and the \ac{OzDES} spectroscopic redshift dataset \citep{OZDES-2015, OZDES-2017}.

\subsection{ATLAS} \label{subsec:atlas}

The \ac{ATLAS} DR3 dataset \citep{franzen} forms the basis for our total catalogue, providing 1.4 GHz radio flux densities on 4780 sources measured using the \ac{ATCA}. The \ac{ATLAS} dataset covers the \ac{ELAIS-S1} and \ac{eCDFS} fields, down to a r.m.s noise level of $\sim15\mu$Jy.

\subsection{SWIRE} \label{subsec:swire}

The \ac{SWIRE} dataset provides infra-red flux densities at 3.6, 4.5, 5.8 and 8.0~$\mu$m, measured using the \ac{SST} \citep{SWIRE}, reaching a $5\sigma$ sensitivity of  5, 9, 43 and 40~$\mu$Jy/beam  respectively. \ac{SWIRE} identifications were found for  4328 radio sources at 3.6~$\mu$m, providing at least a 3.6~$\mu$m flux for $\sim$~91\% of radio sources. All \ac{ATLAS} sources were initially cross matched with \ac{SWIRE} sources, and then the \ac{SWIRE} sources were cross-matched to the \ac{DES} sources.

\subsection{DES} \label{subsec:des}

The \ac{DES} dataset provides \textit{g, r, i} and \textit{z} optical magnitudes (to a depth of \textit{g} = 24.33, \textit{r} = 24.08, \textit{i} = 23.44 and \textit{z} = 22.69), measured using the  Dark Energy Camera mounted on the 4-m Blanco telescope at Cerro Tololo Inter-American Observatory in Chile. \citep{DES}. The \ac{DES} dataset provides optical counterparts for 3102 of our radio sources at \textit{g} band, covering $\sim$65\% of sources.

\subsection{OzDES} \label{subsec:ozdes}

The \ac{OzDES} dataset provides the spectroscopic redshifts required to create any empirical model \citep{OZDES-2015,OZDES-2017}. The spectroscopic redshift master list compiled as the \ac{OzDES} dataset by \citet{OZDES-2017} provides spectroscopic redshifts for 2012 radio sources, covering $\sim$42\% of sources. 

\subsection{Test Sample} \label{subsec:testset}

The \ac{kNN} algorithm works best if the reference data set is shaped such that the feature space is populated homogeneously, i.e. avoiding strong concentrations in a certain region, or sparsely populated regions. In the tests described here, we made no correction for any excess sources in any of the given training samples. Instead, we calculated optical and infrared colours as $\text{c}_i = \text{mag}_i - \text{mag}_{i+1}$. This transformation improves the distribution of the photometric data over the parameter space for each band \citep{norris:2018} and also replaces flux, which is brightness- and redshift-dependent,  by  colour that depends only on the SED \citep{2013MNRAS.428..226P}. We completed this transformation on both the optical magnitudes, and log-transformed infra-red fluxes. Note that this operation reduces the effective number of dimensions of the feature space by 2. 

We  compiled a a full-sensitivity ``DEEP'' sample containing those sources that have photometry at 1.4~GHz, infra-red 3.6, 4.5, 5.8 and 8.0~$\mu$m, optical \textit{g}, \textit{r}, \textit{i} and \textit{z} bands, and a spectroscopic redshift. This provides us with 1408 sources with complete photometry and spectroscopy for our tests. 

We then selected a ``WIDE'' sample to match the depth of photometry expected from the \ac{EMU} survey \citep{EMU}, which will use the SkyMapper survey  which has \textit{r} and \textit{i} limits of $\approx 22$ \citep{SkyMapper} and the AllWISE Infrared Survey \citep{AllWISE} of 3.6~$\mu$m $>$ 26 $\mu$Jy and 4.5~$\mu$m $>$ 56 $\mu$Jy. We  removed the 5.8 and 8.0~$\mu$m data from our sample, and rejected any sources which were undetected at any band at the above limits. The resulting WIDE sample had 760 sources.

As discussed in Section \ref{subsec:atlas}, the \ac{ATLAS} catalogue covers two fields. The \ac{ELAIS-S1} field, which makes up 553 of the 1408 sources in the DEEP data set and 281 of the 760 sources in the WIDE data set, and the \ac{eCDFS} field, which makes up 855 of the 1408 sources in the DEEP dataset and 479 of the 760 sources in the WIDE data set.

\section{Experiments} \label{sec:experiments}
In these experiments, we use the \ac{kNN} regression algorithm to estimate  redshifts, using the following steps:
\begin{enumerate}
\item Compute a distance matrix between all test sources and training sources.
\item Sort the distance matrix by the distance calculated in step 1, identifying the $k$ closest training sources in feature space to the test sources
\item Take the mean redshift of the known sources identified from step 2, and assign it to the test source.
\end{enumerate}

To apply this method, we split our data into a training and test set. Depending on the test being run, the training set was either 70\% of the data set in the case of the full dataset tests (Tests 1-4 in Table \ref{table:tests}), or the entire sub field in the case of the sub-field tests (Tests 5-12 in Table \ref{table:tests}). The remaining 30\% or sub-field was set aside as the test set. 

To avoid under- or over-fitting the data, 10-fold cross validation was used with our training sets - excluding our test sets, minimising the number of incorrect estimates. In all tests, the value of $k$ that minimised the outlier rate varied, and is listed in Table \ref{table:results}.

We  computed 95\% confidence intervals for each redshift prediction using bootstrapping with 1000 iterations.
Bootstrapping is a sampling method that allows us to estimate the variance of sample estimates under the assumption that the population from which the sample is taken is approximately many replications of the sample. The estimated intervals provide the range in which the true redshift is likely to occur, while also providing an indication of the uncertainty of the prediction. These confidence intervals are displayed in the form of error bars in our figures in Section \ref{sec:results}.

In our investigation of the accuracy of \ac{kNN}, we examined the effect of varying the following experimental parameters: 
\begin{itemize}
  \item \textbf{Distance Metric:} We evaluated both the Euclidean distance metric: 
  \begin{equation} 
	\label{eqn:euclidean}  
	d(\vec{p},\vec{q}) = \sqrt{\sum_{i=1}^n (q_i - p_i)^2} 
\end{equation} 
and the Manhattan Taxicab distance metric: 
\begin{equation} 
	\label{eqn:manhattan} 
    d(\vec{p},\vec{q}) = ||\vec{p} - \vec{q}|| = \sum_{i=1}^n |p_i - q_i| 
\end{equation}
Where $p$ and $q$ are vectors containing the features of two sources.
  
  \item \textbf{Depth of Photometry:} We used both the full-sensitivity DEEP sample and the reduced-sensitivity WIDE sample. 
  \item \textbf{Generalisation:} We randomly selected training and test sets from both \ac{ATLAS} sub-fields as one test, and used one \ac{ATLAS} sub-field as the training set and the other as the test set and reverse as additional tests.
\end{itemize}

As these variations are not independent, each needed to be completed in combination with all others, resulting in the 12 experiments listed in Table \ref{table:tests}. 

\section{Results} \label{sec:results}

\begin{figure*}[h!]
\centering
\includegraphics[trim=0 0 0 0, width=.23\textwidth]{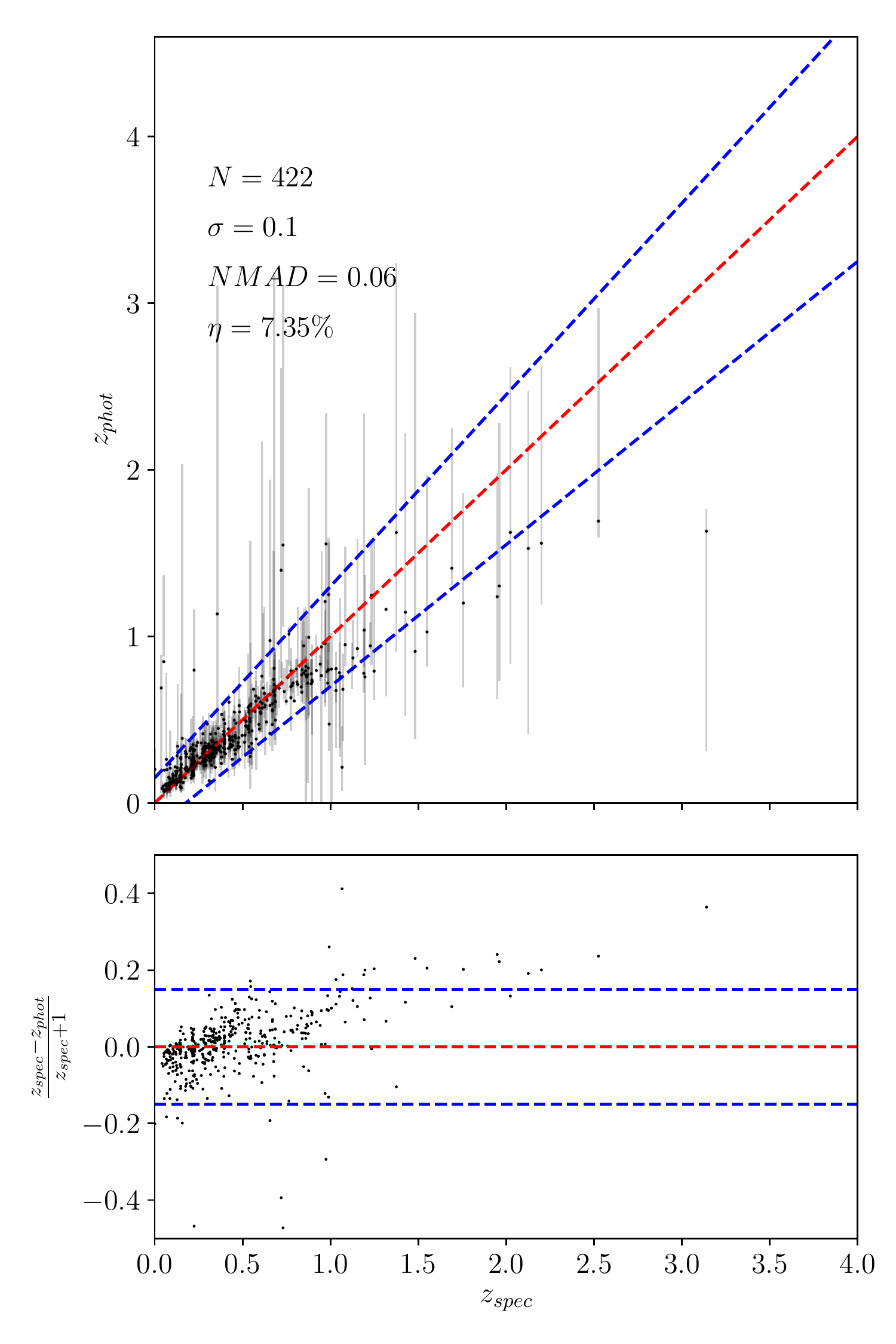}
\includegraphics[trim=0 0 0 0, width=.23\textwidth]{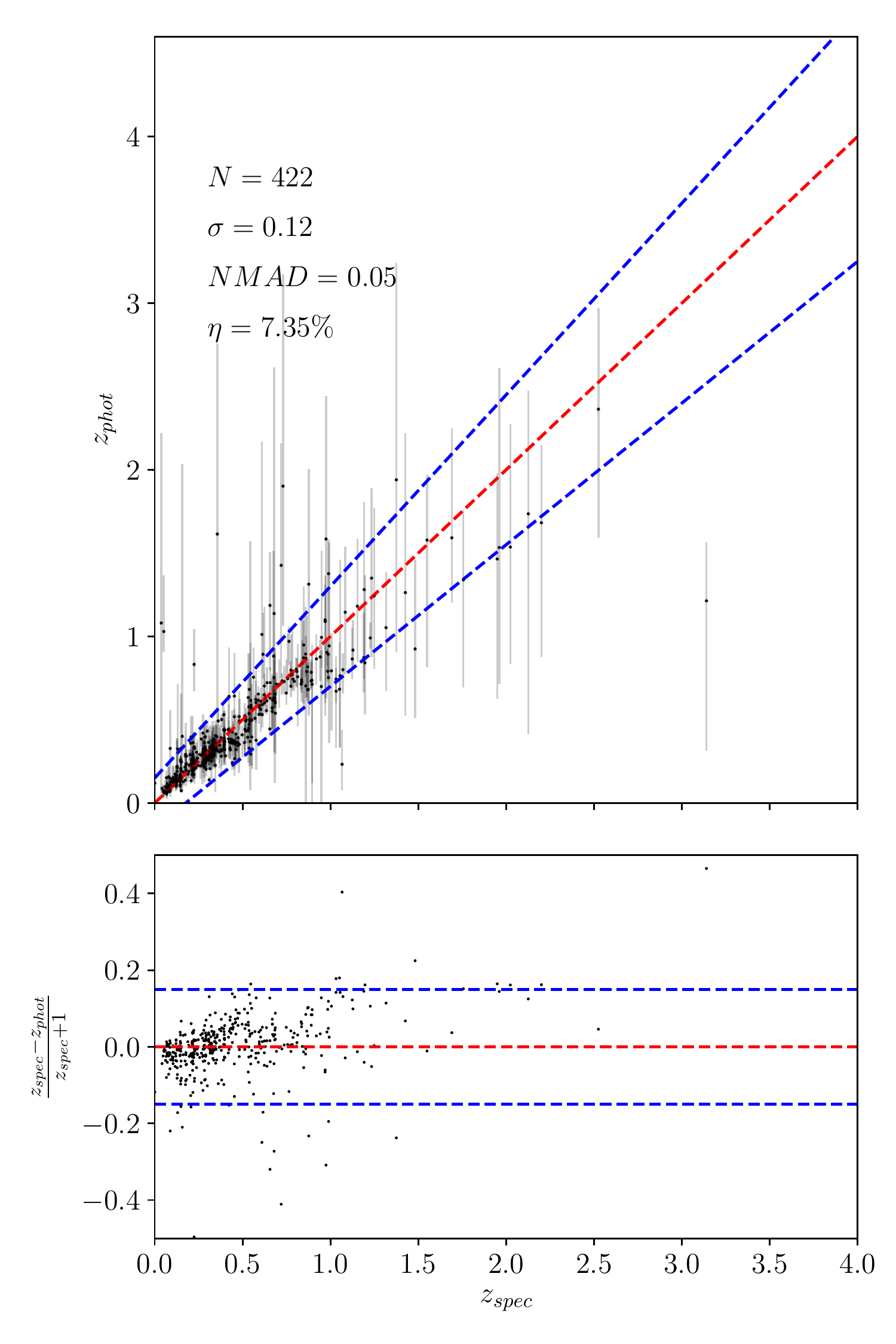}
\includegraphics[trim=0 0 0 0, width=.23\textwidth]{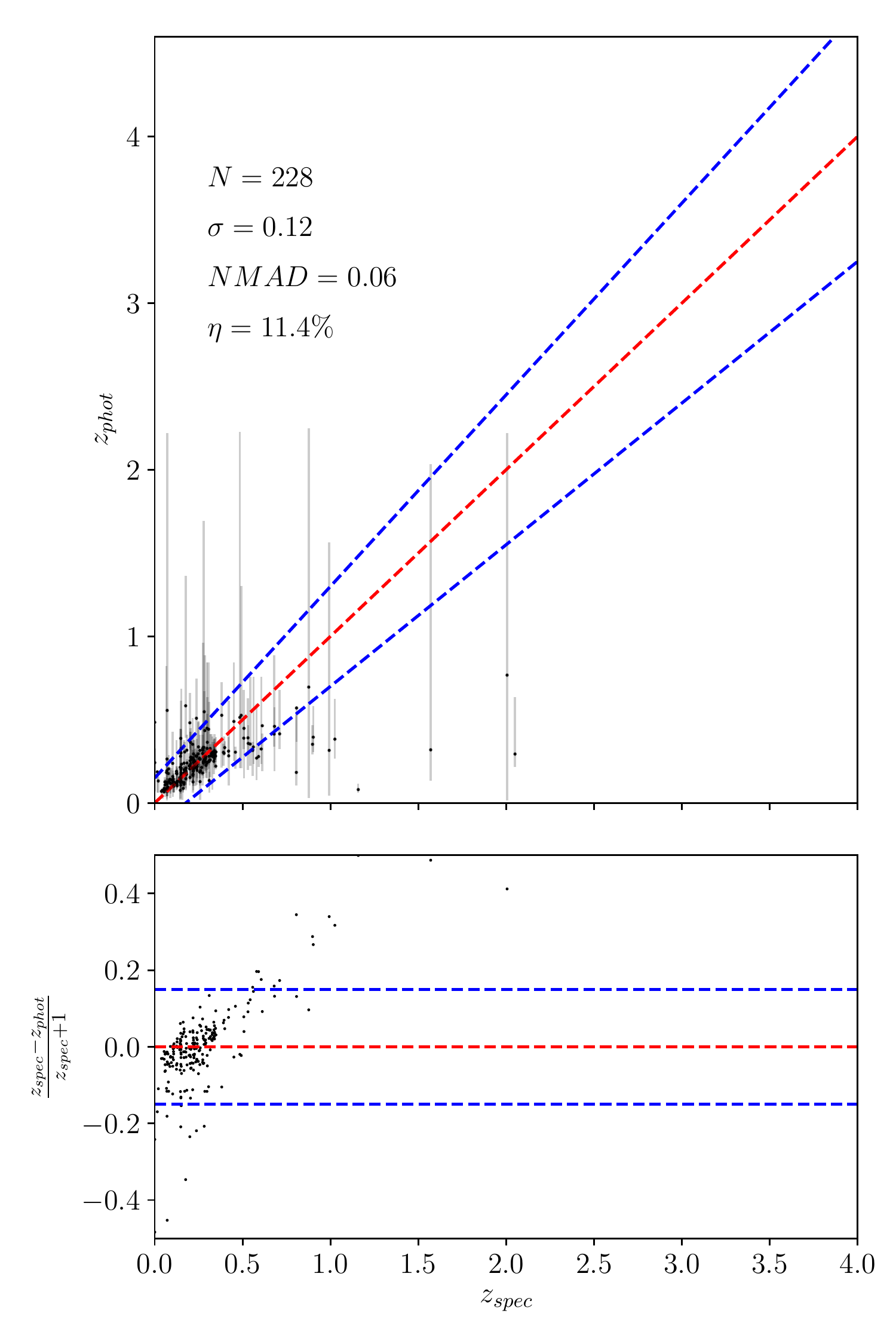}
\includegraphics[trim=0 0 0 0, width=.23\textwidth]{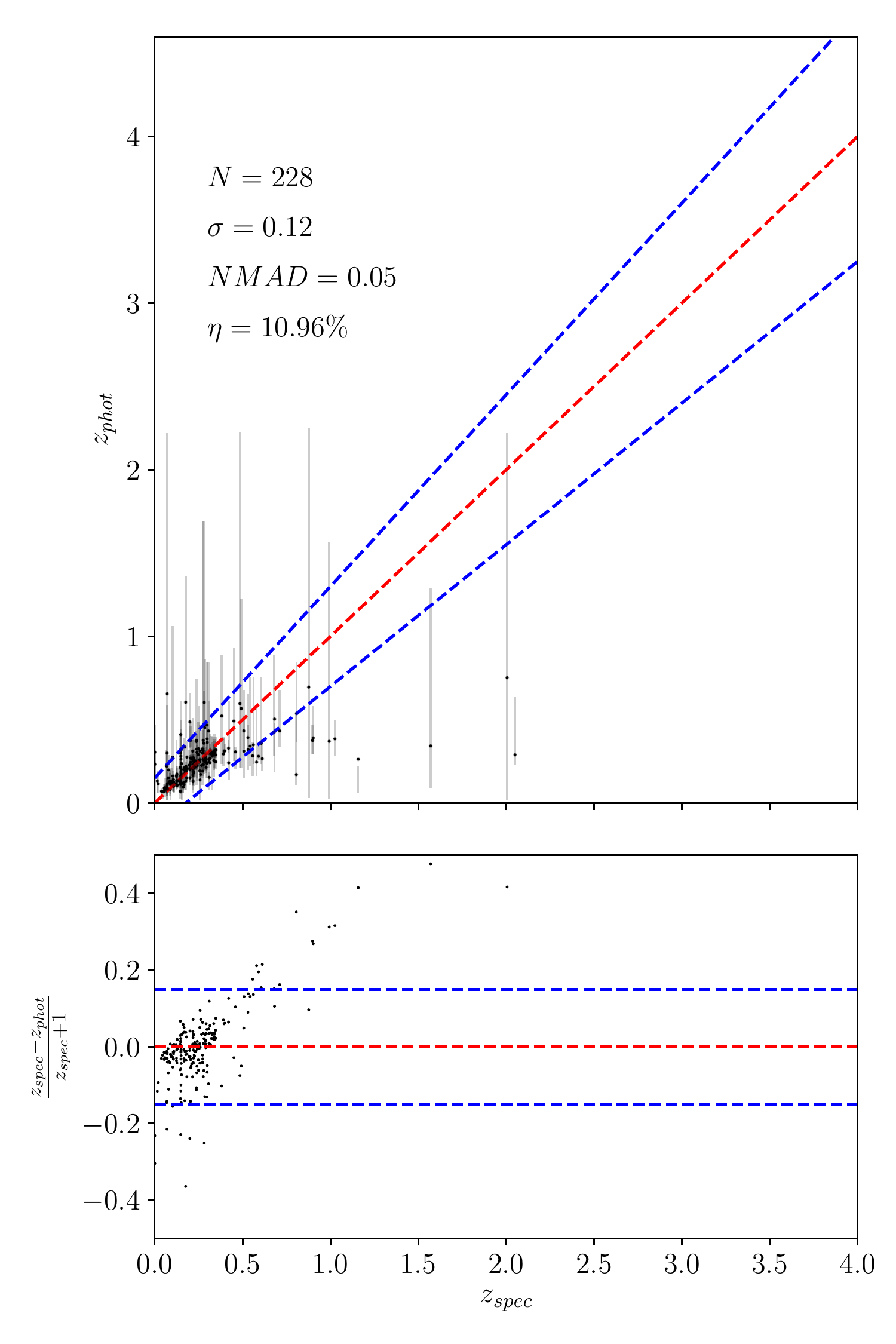}
\caption{Summary of the results from Tests 1-4, from left to right. All results displayed here have been trained on sources randomly sampled from the entire \ac{ATLAS} data set. The top panels show the distribution of $z_{spec}$ vs $z_{photo}$, and the lower panels show the normalised residuals vs the $z_{spec}$. The dashed red line represents $z_{spec} = z_{photo}$, and the dashed blue lines represent the outlier boundary, calculated using Equation \ref{eqn:outlier}.}
\label{fig:test1.4}
\end{figure*}

\begin{figure*}[h!]
\centering
\includegraphics[trim=0 0 0 0, width=.23\textwidth]{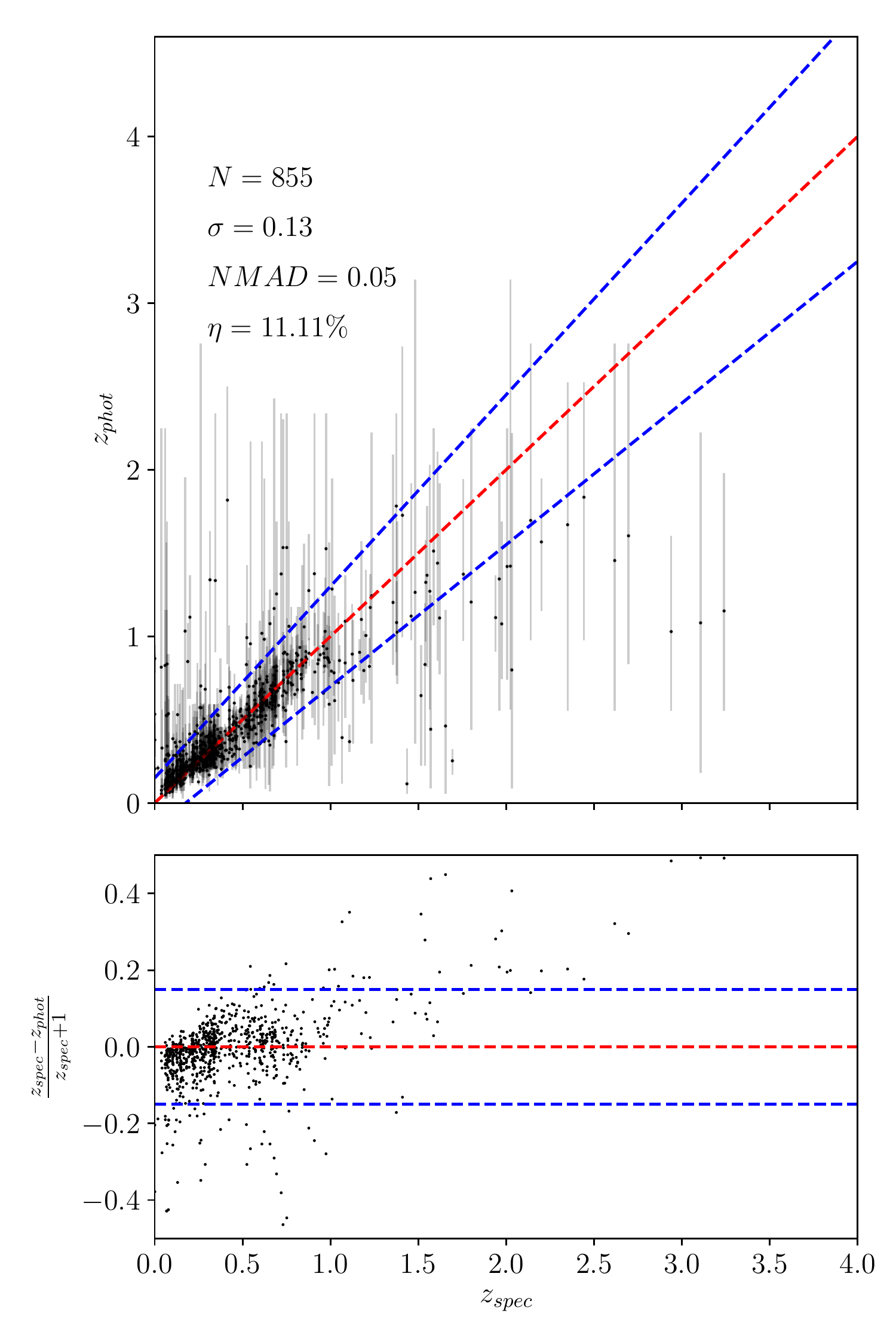}
\includegraphics[trim=0 0 0 0, width=.23\textwidth]{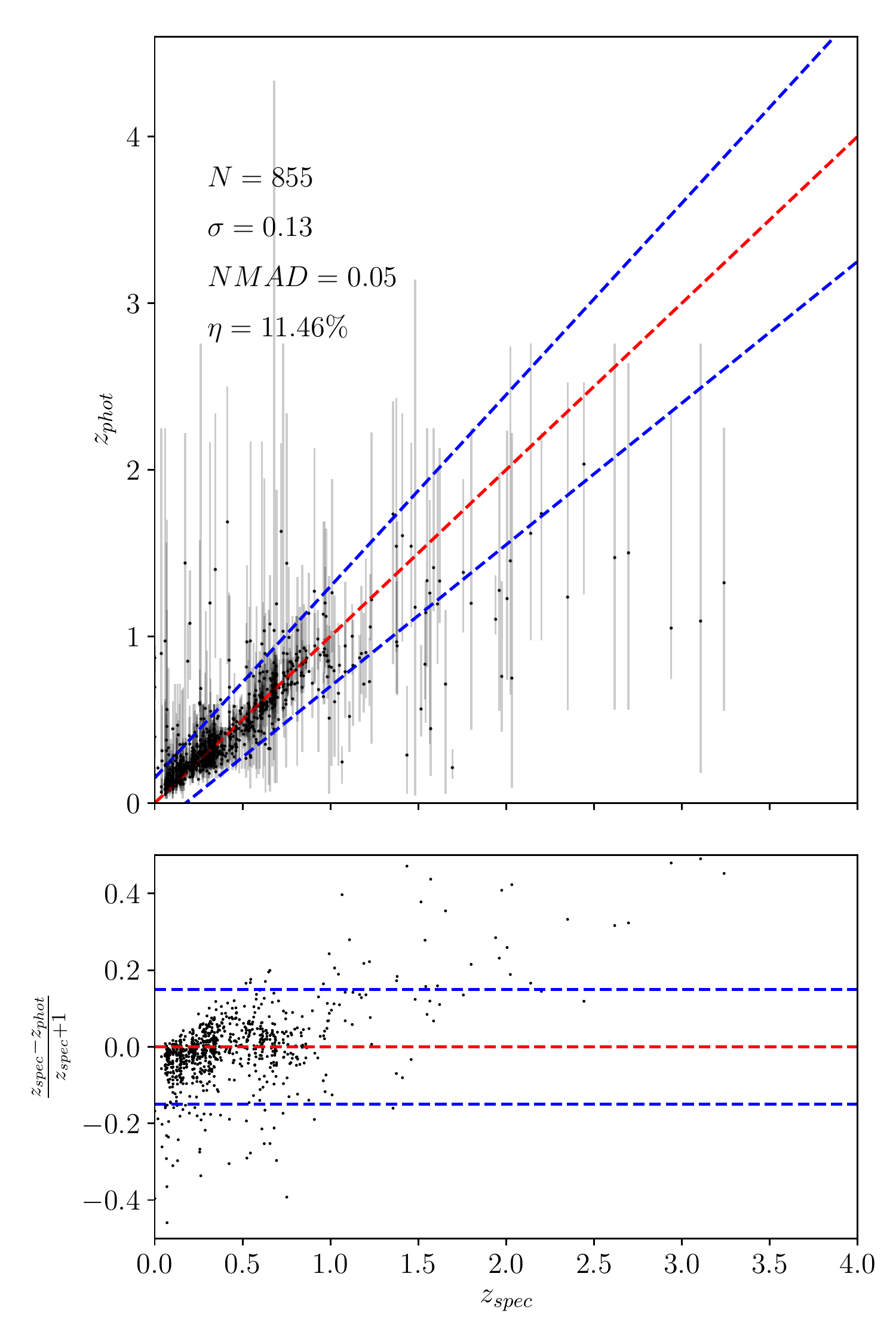}
\includegraphics[trim=0 0 0 0, width=.23\textwidth]{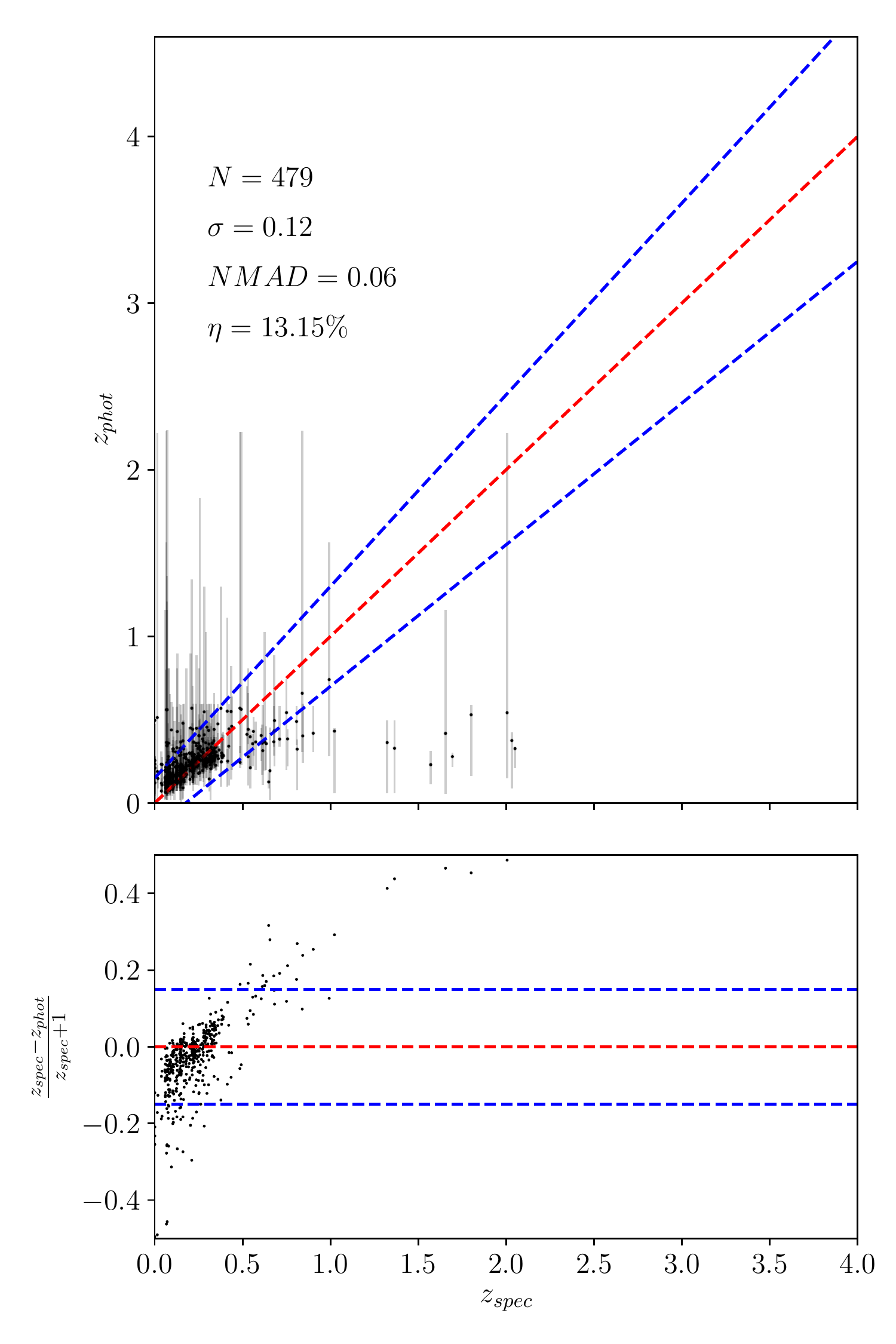}
\includegraphics[trim=0 0 0 0, width=.23\textwidth]{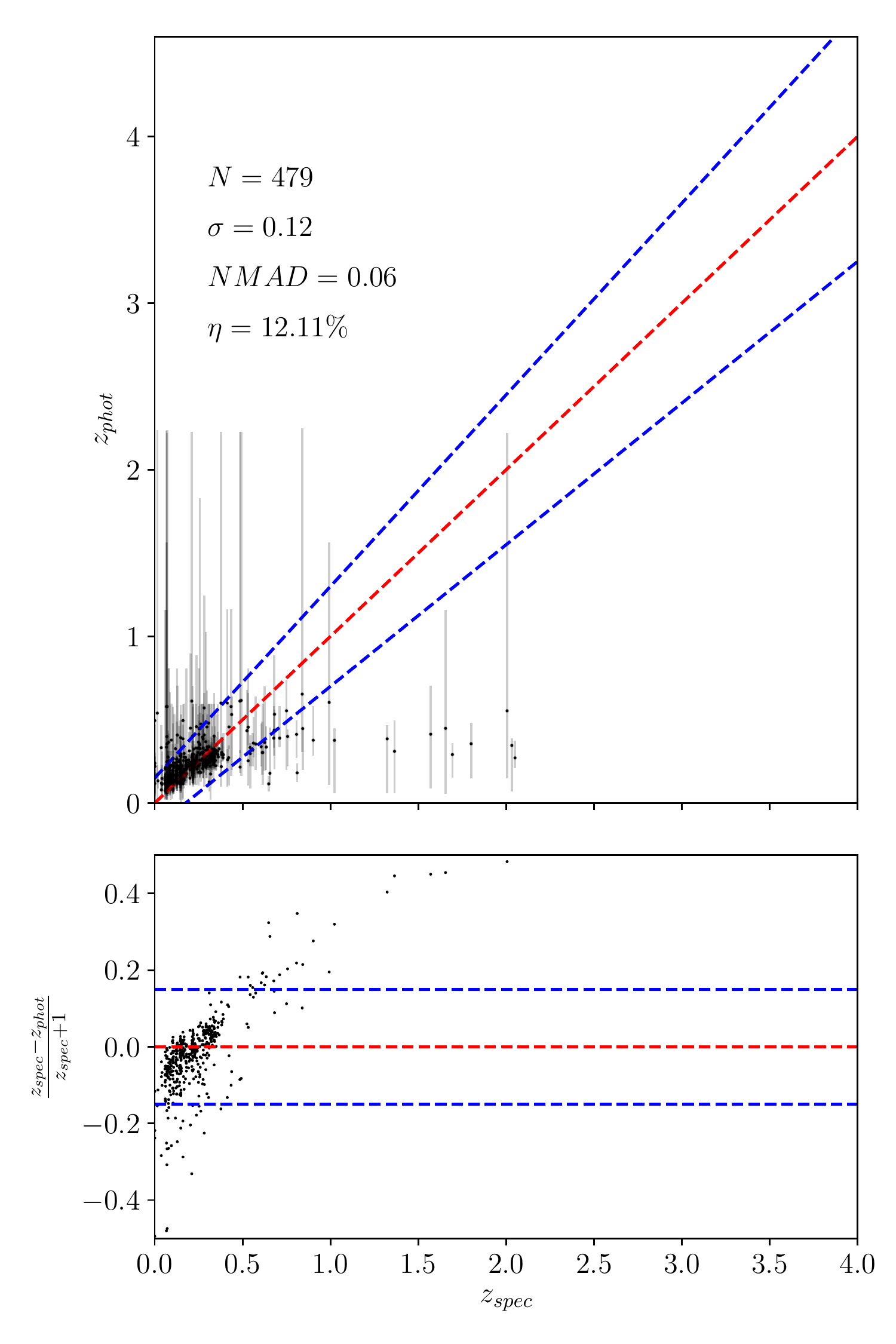}
\caption{Summary of the results from Tests 5-8, from left to right. All results displayed here have been trained on sources found exclusively in the \ac{ELAIS-S1} field. Other details as in Figure \ref{fig:test1.4}.}
\label{fig:test5.8}
\end{figure*}

\begin{figure*}[h!]
\centering
\includegraphics[trim=0 0 0 0, width=.23\textwidth]{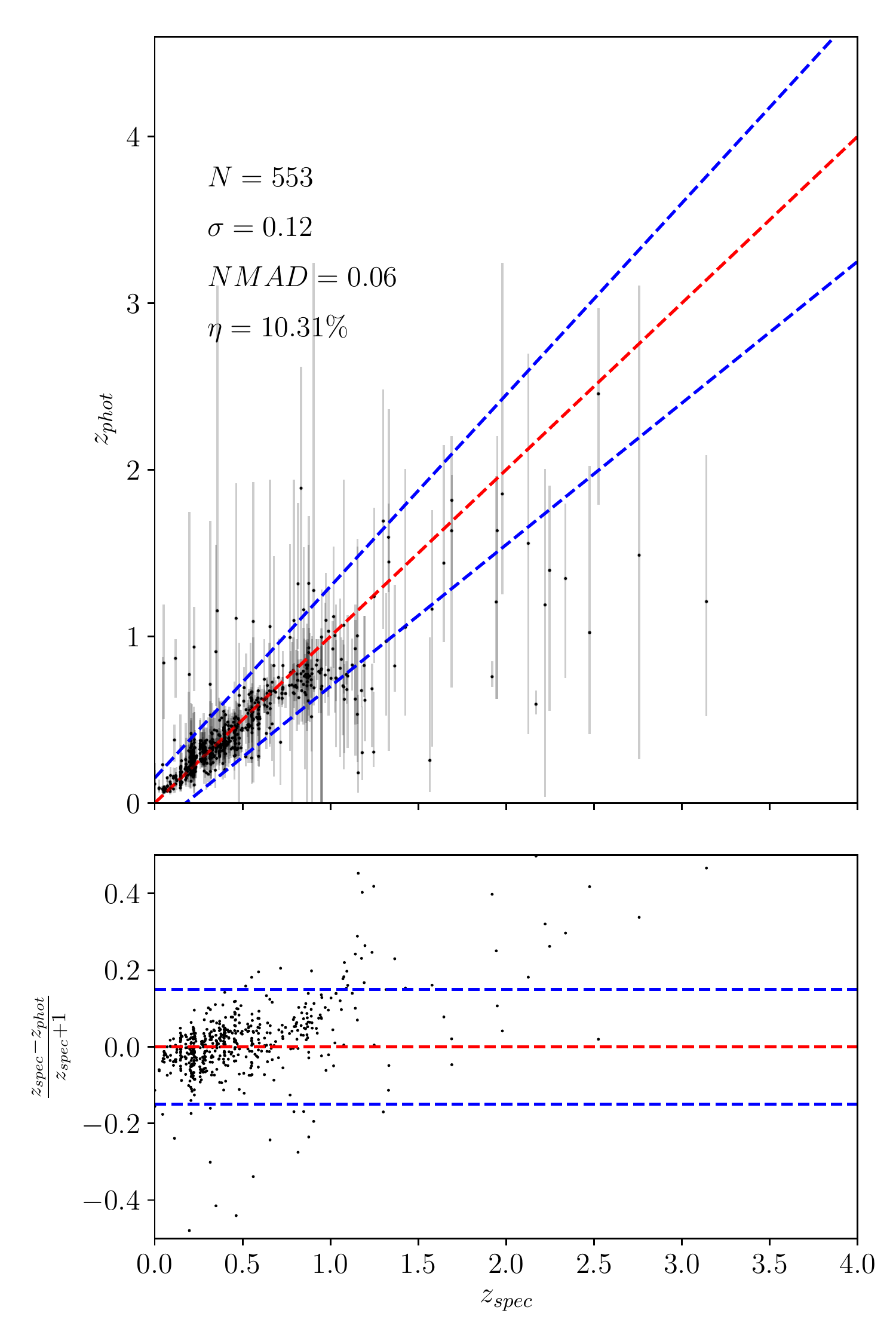}
\includegraphics[trim=0 0 0 0, width=.23\textwidth]{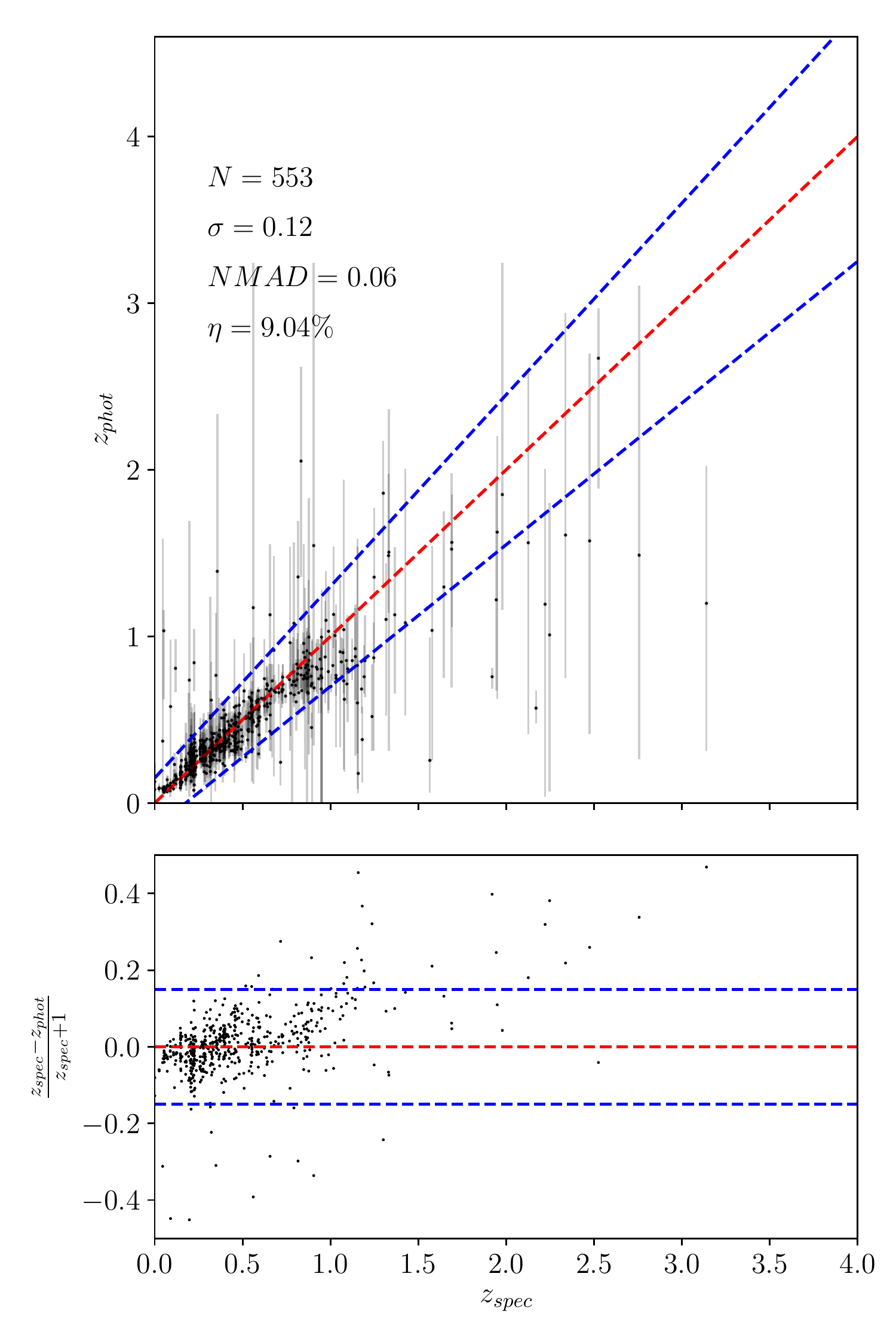}
\includegraphics[trim=0 0 0 0, width=.23\textwidth]{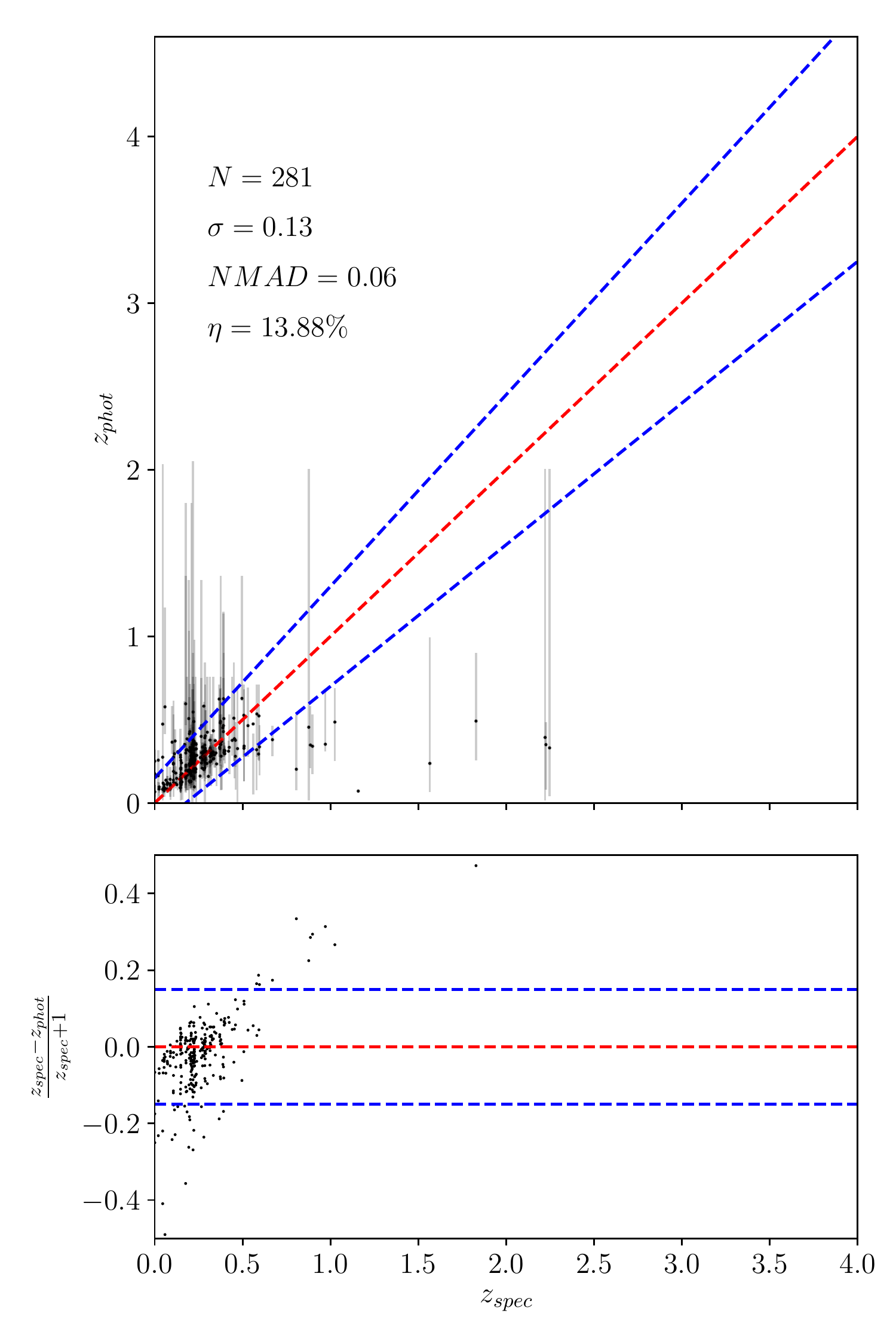}
\includegraphics[trim=0 0 0 0, width=.23\textwidth]{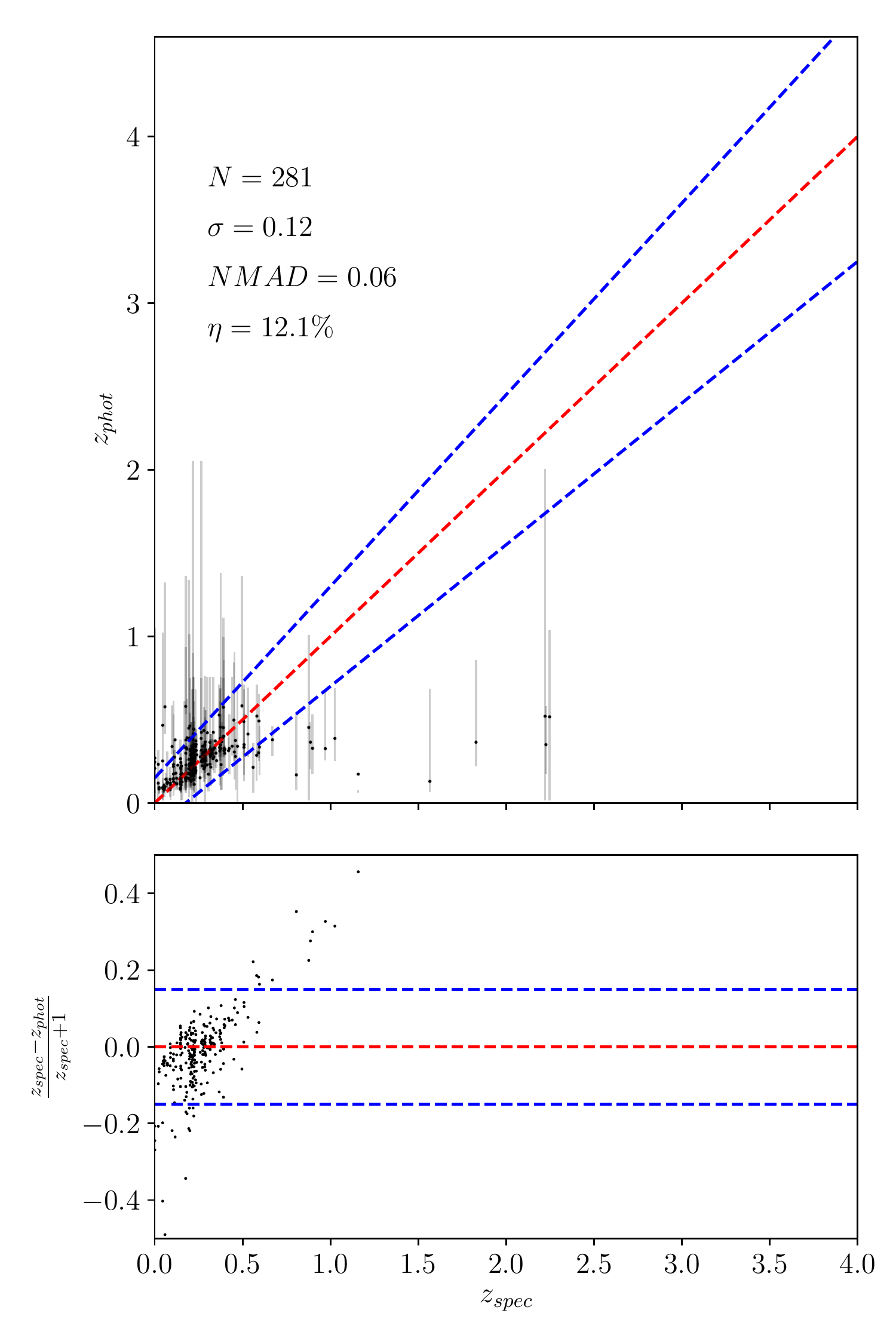}
\caption{Summary of the results from Tests 9-12, from left to right.All results displayed here have been trained on sources found exclusively in the \ac{eCDFS} field. Other details as in Figure \ref{fig:test1.4}.}
\label{fig:test9.12}
\end{figure*}

We present the results of the experiments in Table \ref{table:results} and Figures \ref{fig:test1.4} to \ref{fig:test9.12}. 

In Table \ref{table:results} and Figures \ref{fig:test1.4} to \ref{fig:test9.12}, we calculate the outlier rate $\eta$ as:
\begin{equation} 
	\label{eqn:outlier} 
    \eta = \frac{\text{count}(|\Delta z| > 0.15 \times (1 + z_{spec}) )}{\text{Number Of Sources}} \times 100
\end{equation}
where $\Delta z = z_{spec} - z_{photo}$ and the normalised median absolute deviation $\sigma_{\text{NMAD}}$ as: 
\begin{equation} 
	\label{eqn:nmad} 
    \sigma_{\text{NMAD}} = 1.4826 \times \text{median}( |X_i - \text{median}(X) |)
\end{equation}
where $X$ is a vector of residuals

\begin{table*}[h!]
  \begin{center}
    \caption{Details of experiments completed, including the experiment number, training and test set sizes, the distance metric used, the dataset used and where the training sample came from.}
    \label{table:tests}
    \begin{tabular}{ c c c c c c }
       Experiment  & Size of & Size of& Distance  & Data set  & Training  \\ 
       Number & Training Set & Test Set & Metric & Used & Sample \\
       \hline 
       1 & 986 & 422 & Manhattan & DEEP & Random \\
       2 & 986 & 422 & Euclidean & DEEP & Random \\
       3 & 532 & 228 & Manhattan & WIDE & Random \\
       4 & 532 & 228 & Euclidean & WIDE & Random \\
       5 & 553 & 855 & Manhattan & DEEP & \ac{ELAIS-S1} \\
       6 & 553 & 855 & Euclidean & DEEP & \ac{ELAIS-S1} \\
       7 & 281 & 479 & Manhattan & WIDE & \ac{ELAIS-S1} \\
       8 & 281 & 479 & Euclidean & WIDE & \ac{ELAIS-S1} \\
       9 & 855 & 553 & Manhattan & DEEP & \ac{eCDFS} \\
       10 & 855 & 553 & Euclidean & DEEP & \ac{eCDFS} \\
       11 & 479 & 281 & Manhattan & WIDE & \ac{eCDFS} \\
       12 & 479 & 281 & Euclidean & WIDE & \ac{eCDFS} \\
     \end{tabular}
  \end{center}
\end{table*}

In Figures \ref{fig:test1.4}, \ref{fig:test5.8} and \ref{fig:test9.12},  the top panels show the distribution of $z_{spec}$ vs $z_{photo}$, and the lower panels show the normalised residuals vs the $z_{spec}$, where the normalised residuals are calculated by:
\[ \frac{\Delta z}{z_{spec} + 1} \]

For every source plotted on the top panels, we have provided error bars representing the 95\% confidence interval of each redshift, calculated using the bootstrap method. On each of the plots, we display multiple statistics:
\begin{itemize}
\item $N$ - The number of sources in the test sample
\item $\sigma$ - Standard deviation of the residual error, calculated typically 
\item $\text{NMAD}$ - Standard deviation of the residual error, calculated using the normalised absolute deviation (Equation \ref{eqn:nmad})
\item $\eta$ - Outlier rate, calculated using Equation \ref{eqn:outlier}
\end{itemize}

These results have all been summarised in Table \ref{table:results}.

\begin{table}[h!]
  \begin{center}
    \caption{Summary of the results from all tests. We include the experiment number, the test size, standard deviation calculated typically and by normalised absolute deviation and outlier rate.}
    \label{table:results}
    \begin{tabular}{ c c c c c c }
       Experiment & Test & Best & Standard & NMAD & Outlier  \\ 
       Number & Size & $k$ & Deviation &  & Rate \\
       \hline 
       1 & 422 & 3 & 0.1 & 0.06 & 7.35\% \\
       2 & 422 & 14 & 0.12 & 0.05 & 7.35\% \\
       3 & 228 & 3 & 0.12 & 0.06 & 11.40\% \\
       4 & 228 & 2 & 0.12 & 0.05 & 10.96\% \\
       5 & 855 & 13 & 0.13 & 0.05 & 11.11\% \\
       6 & 855 & 9 & 0.13 & 0.05 & 11.46\% \\
       7 & 479 & 3 & 0.12 & 0.06 & 13.15\% \\
       8 & 479 & 4 & 0.12 & 0.06 & 12.11\% \\
       9 & 553 & 4 & 0.12 & 0.06 & 10.31\% \\
       10 & 553 & 3 & 0.12 & 0.06 & 9.04\% \\
       11 & 281 & 2 & 0.13 & 0.06 & 13.88\% \\
       12 & 281 & 2 & 0.12 & 0.06 & 12.10\% \\
     \end{tabular}
  \end{center}
\end{table}

\subsection{Distance Metric}
\label{subsec:resultsDistMetric}

Across all of our experiments, we have found negligible difference between using the Manhattan Taxicab distance metric and the Euclidean Distance metric. This indicates that there were few significant outliers when finding the $k$ nearest neighbours, as the Manhattan Taxicab distance will minimise the effect of outliers. In Tests 3 vs 4, 7 vs 8, and 11 vs 12, we find that Euclidean Distance provides a slightly lower outlier rate. In Test 5 vs 6 and 9 vs 10, Manhattan Taxicab distance provides the better option with Test 1 vs 2 being equal. In no case however, does one distance metric have a difference in outlier rate greater than 1.78\%. 

\subsection{Depth of Field}
\label{subsec:resultsDepth}

In all cases, we find that the outlier rate is higher in the WIDE dataset when compared with the DEEP dataset, as shown by the right-hand pair of panels in Figures~\ref{fig:test1.4} to \ref{fig:test9.12}. For the \ac{kNN} regression algorithm, this is expected in current catalogues. In the process of modifying the DEEP dataset to become the WIDE dataset, all sources that we removed are at the fainter end of the dataset, which are typically the high redshift sources. This leaves the WIDE dataset with a more heavily positively skewed distribution of $z_{spec}$, with the majority of sources being below $z = 0.5$. For the sources at high redshifts, the \ac{kNN} algorithm tends to fail because of the paucity of high-redshift sources, forcing low-redshift sources into the group of nearest neighbours.  

A better test would use a larger data set, with a larger population of high-redshift sources, but such a dataset is not yet available.

\subsection{Generalisation}
\label{subsec:resultsGeneralisation}

While the best-case DEEP experiments using a random sample as the training set attain the best results, experiments that train on one ATLAS field, and test on a different ATLAS field, are not much worse. We can attribute the better results in the former case to a more uniform and statistically consistent training sample. This indicates that, with a more consistent training sample, the kNN regression algorithm should perform well across different sections of sky.

\section{Implications for large-scale radio surveys}
\label{sec:implications}

Table \ref{table:results} shows that redshifts can be measured to an accuracy of about 6\% (NMAD) to 12\% (standard deviation), with an outlier rate of about 11\%, and this result remains true for all sources for which photometry is available, independently of the depth of the photometry.

If we assume that the DEEP sample has the same radio sensitivity as EMU, and that the WIDE sample has the same optical/infra-red photometric depth as that available for EMU sources, then the relative sizes of the DEEP and WIDE samples implies that $\sim$ 45\% of EMU sources will have the required photometry for redshift measurement.  

We can therefore conclude, based on these tests, that about 40\% of EMU sources will have redshifts available, or a total of about 28 million radio sources.

We plan to extend this work by investigating the effect of (a) using the non-detection information, (b) using a more sophisticated metric that allows missing values and measurement bounds, (c) carefully modelling the sensitivity limits of the available photometric surveys, (d) incorporating other data types (such as radio and X-ray), (e) quantising redshift to provide a classification problem rather than a regression problem, (f) obtaining more high-redshift training data from deep surveys in small fields. Future work will continue in this direction.

\section{Conclusion} \label{sec:conclusion}

The main result from these preliminary experiments is that, using the kNN algorithm, we can make good estimates of redshifts in large radio surveys given the photometry that is likely to be available, although further work is expected to improve results further. Around 90\% of EMU sources with optical/infra-red photometry will have a reliable estimated redshift, resulting in redshifts for $\sim$ 40\% of EMU sources. However, we expect that future work will result in an even higher fraction of sources with useful redshifts.

We found no obvious difference in the results provided by Manhattan Taxicab distance and Euclidean Distance. However, greater benefits may be obtained from self-learned distance metrics that can weight features based on their relevance, and can deal with missing values.

We found that the DEEP dataset reported better results than the WIDE dataset, probably because the WIDE survey contains mainly low-redshift sources with the few high-redshift sources having to be estimated from low-redshift sources. Further work will characterise what fraction of sources have the required photometry at higher redshifts.

We found that there was no obvious difference in success rate if the algorithm used training and test sets from spatially separate fields in the sky. Experiments 7-8 and 11-12 (different field training sets on the WIDE dataset) suggest that we can overcome the lack of high redshift sources in training sets by drawing training sets from small, deep fields and applying them to targets covering the entire sky.

Finally, this paper has demonstrated that with sufficient redshift coverage in the training set, the \ac{kNN} algorithm provides acceptable error rates when estimating the redshift of radio galaxies.

\begin{acronym}[ELAIS-S1]
\acro{ATCA}{Australia Telescope Compact Array}
\acro{ATLAS}{Australia Telescope Large Area Survey}
\acro{DES}{Dark Energy Survey}
\acro{eCDFS}{extended Chandra Deep Field South}
\acro{ELAIS-S1}{European Large Area ISO Survey--South 1}
\acro{EMU}{Evolutionary Map of the Universe}
\acro{kNN}{$k$-Nearest Neighbours}
\acro{OzDES}{Australian Dark Energy Survey}
\acro{SDSS}{Sloan Digital Sky Survey}
\acro{SST}{Spitzer Space Telescope}
\acro{SWIRE}{Spitzer Wide-Area Infrared Extragalactic Survey}
\end{acronym}

\section*{Acknowledgements}
We thank Stefano Cavuoti and his colleagues for use of the code to plot our results.

\bibliographystyle{aasjournal}
\bibliography{kNN_Redshift}



\end{document}